\pdfoutput=1
\documentclass[runningheads]{llncs}

\usepackage[
  left=3cm,   
  right=3cm,
  top=3cm,
  bottom=3.0cm
]{geometry}
\usepackage[T1]{fontenc}
\usepackage{comment}
\usepackage{amssymb}
\usepackage[usenames,dvipsnames]{xcolor}
\usepackage{listings}
\usepackage[numbers]{natbib}
\lstdefinelanguage{Julia}%
  {morekeywords={abstract,break,case,catch,const,continue,do,else,elseif,%
      end,export,false,for,function,immutable,import,importall,if,in,%
      macro,module,otherwise,quote,return,switch,true,try,type,typealias,%
      using,while},%
   sensitive=true,%
   breaklines=-true,%
   alsoother={\$},%
   morecomment=[l]\#,%
   morecomment=[n]{\#=}{=\#},%
   morestring=[s]{"}{"},%
   morestring=[m]{'}{'},%
   frame=single,%
   numbers=left,%
}[keywords,comments,strings]%

\usepackage{graphicx}
\begin{document}

\title{Toward Portable GPU Performance: Julia Recursive Implementation of TRMM and TRSM}

%
%

\makeatletter
\newcommand{\printfnsymbol}[1]{%
  \textsuperscript{\@fnsymbol{#1}}%
}
\makeatother

\author{
    Submitted to the \emph{Workshop on Asynchronous Many‑Task Systems and Applications}
          (WAMTA’25) on 02/07/2025 — \url{https://wamta25.github.io/}. [In press] This is a preprint version.
        \\
    \vspace{.15in}
    Vicki Carrica\thanks{equal contribution}\and
    Maxwell Onyango\printfnsymbol{1} \and
    Rabab Alomairy\and 
    Evelyne Ringoot \and \\
    James Schloss \and
    Alan Edelman
}

\authorrunning{Carrica et al.}
\titlerunning{Performance Portable TRMM and TRSM}
\institute{
    Computer Science \& Artificial Intelligence Laboratory, \protect \\Massachusetts Institute of Technology, USA. \\
    \email{vickicar@mit.edu, maxmlewa@mit.edu, rabab.alomairy@mit.edu,  eringoot@mit.edu, \\ jars@mit.edu, edelman@mit.edu} 
}

%
\maketitle              
\begin{abstract}
This paper presents a performant and portable recursive implementation of triangular matrix-matrix multiplication (TRMM) and triangular solve (TRSM) operations in Julia for GPUs, which form  
 the backbone of many other linear algebra algorithms. This work is based on an existing recursive implementation for TRMM and TRSM, which restructures the operations to include general matrix-matrix multiplication (GEMM) calls, facilitating better
utilization of the GPU memory hierarchy, and reducing latency overhead. The unified implementation in Julia harnesses the language's multiple-dispatch and metaprogramming capabilities through the existing GPUArrays and KernelAbstractions frameworks, enabling performant hardware-agnostic execution across different GPU architectures. By supporting a consistent API, this implementation allows users to seamlessly switch between different GPU backends. The recursive hardware-agnostic implementation we present achieves performance comparable to vendor-optimized (cuBLAS/rocBLAS) libraries for larger matrix sizes and provides such methods for the first time to Apple Silicion hardware with  only a few hundred lines of code, demonstrating the power of unified implementations.



\end{abstract}

\section{Introduction}
Triangular matrix-matrix multiplication (TRMM) and triangular solve (TRSM) are foundational operations in dense linear algebra and part of the BLAS Level 3 (BLAS3) specification, which encompasses high-performance routines designed to operate on blocks of data ~\cite{gates2025evolution}. These operations underpin numerous scientific computations and engineering applications. TRMM computes the product of a triangular matrix with another matrix, facilitating efficient transformations and updates to the matrix data. TRSM solves triangular systems of linear equations, a critical step in algorithms for matrix inversion, LU decomposition, and other matrix factorizations.~\cite{blackford2002updated}



Existing implementations for TRMM and TRSM on GPU hardware, such as in NVIDIA cuBLAS, often fail to achieve peak performance comparable to general matrix-matrix multiplication (GEMM), necessitating novel approaches to optimize data reuse and mitigate latency~\cite{charara2016redesigning, charara2017framework}. 
This is in-part because the triangular structure introduces challenges such as Write-After-Read (WAR) and Read-After-Write (RAW) dependencies, which limit parallelism and incur excessive memory traffic on GPUs. 
This has encouraged the development of recursive formulations that decompose these operations into General Matrix Multiply (GEMM) calls interspersed with small triangular updates.
This restructuring as a series of recursive kernel launches reduces memory accesses, maximizes concurrency, and aligns better with GPU memory hierarchies~\cite{charara2016redesigning}.
For example, the KBLAS library \cite{charara2016redesigning} employs recursive algorithms to improve TRMM and TRSM performance by leveraging GEMM's parallelism and optimized memory access patterns.
These enhancements can result in speedups of up to eightfold for large matrices compared to state-of-the-art libraries.







Several high-performance linear algebra libraries have long provided GPU-based implementations of tiled triangular matrix solve (TRSM) and triangular matrix-matrix multiplication (TRMM) as part of their numerical computing frameworks. Notable among these are SLATE~\cite{gates2025evolution, alomairy2022communication}, which offers a modern distributed dense linear algebra interface optimized for heterogeneous architectures; Chameleon~\cite{faverge2023programming}, which leverages the StarPU runtime system to efficiently schedule tasks across CPUs and GPUs; and DPLASMA~\cite{hoque2017dynamic}, which builds on the PaRSEC runtime~\cite{alomairy2022high} for scalable distributed-memory execution of dense linear algebra workloads. These libraries have demonstrated the effectiveness of task-based parallelism in optimizing tiled TRSM and TRMM computations by exposing fine-grained parallelism and ensuring efficient data movement across heterogeneous architectures.

In this work, we demonstrate how modern software abstractions for GPU programming models can effectively characterize classic linear algebra algorithms based on triangular matrix operations.
Our approach offers extensive hardware portability and software flexibility while incurring minimal performance overhead.
It is also the first implementation of such methods on Apple Silicon devices.
Our main contributions are as follows:

\begin{itemize} 
    \item A unified, accelerated implementation of TRMM and TRSM that extends existing methods and, for the first time, supports multiple platforms, including Apple Silicon. The code can be found in ~\cite{the-code}.
    \item A performance-optimized design that achieves throughput comparable to state-of-the-art libraries such as cuSOLVER and rocBLAS.
    \item A case study illustrating how a unified abstraction framework can enable scalable, portable solutions for GPU-accelerated functions covering most GPU vendors (AMD, NVIDIA, Apple Silicon) and parallel CPU execution.
\end{itemize}

Overall, we emphasize the balance between portability and efficiency offered by modern GPU software abstractions and present this approach as a modern solution for future challenges in heterogeneous computing systems.
This work serves as an important first step towards the development of a truly hardware independent and performance portable linear algebra system in Julia.
This paper is organized as follows:
we discuss necessary background information in Section~\ref{sec:bg},implementation details in Section~\ref{sec:imp}, performance results in Section~\ref{sec:res}, and conclude in Section~\ref{sec:con}. 


\label{sec:intro}

\section{Background}

This section provides the necessary background on TRSM and TRMM operations, the abstraction framework in Julia for high-performance computing (HPC) that facilitate portable and efficient GPU programming from which we benefit in this work.

\subsection{TRMM and TRSM Operations}

The TRMM algorithm computes a matrix-matrix product where one input matrix is triangular. The operation is defined as:

\begin{equation}
   C := \alpha*op(A)*B
\end{equation}
or
\begin{equation}
    C := \alpha*B*op(A),
\end{equation}
where $A \in \mathbb{R}^{n\times n}$ is an upper or lower triangular matrix, op(A) is one of $op(A) = A$, or $op(A) = A^T$, or $op(A) = conjg(A^T)$, and $B \in \mathbb{R}^{n\times m}$ is a dense matrix. 

 The TRSM algorithm solves a triangular matrix equation for the matrix $X$. The operation is defined as:
 \begin{equation}
     op(A)*X = \alpha*B
 \end{equation}
 or alternatively the system
 \begin{equation}
     X*op(A) = \alpha*B,
 \end{equation}
 where $X \in \mathbb{R}^{n\times m}$,   $A \in \mathbb{R}^{n\times n}$ ia unit, or non-unit, upper or lower triangular matrix, op(A) is one of $op(A) = A$, or $op(A) = A^T$, or $op(A) = conjg(A^T)$, $B \in \mathbb{R}^{n\times m}$ is a dense matrix, and alpha is a scalar.

\subsection{Abstraction framework in Julia for code efficiency}

In this work, we provide performance portable implementations of the recursive TRMM and TRSM methods.
Historically, each hardware vendor provides their own programming interface and linear algebra library for such operations (e.g. CUDA and cuBLAS for NVIDIA or ROCm and rocBLAS for AMD hardware).
These libraries often follow different  approaches, posing challenges in heterogeneous computing environments when users have a variety of different hardware available.
To address this issue, several languages have been created that can be both performant and portable to different hardware (e.g. OpenCL, SyCL, Kokkos\cite{davis2024evaluative}, mojo, and Julia \cite{churavy2022bridging, churavy2024}).

Julia language provides core abstractions that enable the flexible creation of linear algebra routines capable of executing on diverse hardware platforms~\cite{dla}. It manages concurrency using a task-based model, where tasks are defined at the program level and scheduled onto hardware or OS threads by Julia’s runtime. Julia features a ``just-barely-ahead-of-time'' compiler that will compile user code only when types can be inferred and statically known.
For CPU execution, Julia will compile to the LLVM (Lower Level Virtual Machine) intermediate representation to achieve equivalent performance to other LLVM languages such as C and Rust, while also continuing to provide features expected from higher level languages like R, MATLAB, and Python.
In the case of GPU execution, the {\tt GPUCompiler.jl} and {\tt SPIRV.jl} packages will emit lower level code to an LLVM-like intermediate representation that matches the appropriate hardware (e.g. NVPTX for CUDA and NVIDIA hardware).
This approach directly contrasts languages like mojo that compile down to another Multi-Level Intermediate Representation (MLIR) before code lowering to LLVM dialects.

This work leverages the {\tt GPUArrays.jl} and {\tt KernelAbstractions.jl} packages for performance portability.
{\tt GPUArrays.jl} is a suite of tools that allow Julia users to efficiently generate GPU code for different hardware vendors by modifying the type signature of the input data.
This package also holds specialized routines to provide a base-level of performance to all GPU users and allows users to perform array-level abstractions such as broadcasting.
{\tt KernelAbstractions.jl} is a kernel-level interface that is supported by different hardware backends ({\tt CUDA.jl}~\cite{besard2019}, {\tt AMDGPU.jl}~\cite{AMDGPU}, {\tt oneAPI.jl}~\cite{oneapi}, {\tt Metal.jl}~\cite{metal} and parallel CPU execution.
With these tools available, we have condensed much of the functionality of a large library (KBLAS \cite{charara2016redesigning, charara2017framework}) into a few hundred lines of code~\cite{the-code}.

\label{sec:bg}

\section{Implementation} 
\label{sec:imp}

\subsection{Recursive Framework}

\begin{figure}[th]
	\centering
	\includegraphics[scale=0.5]{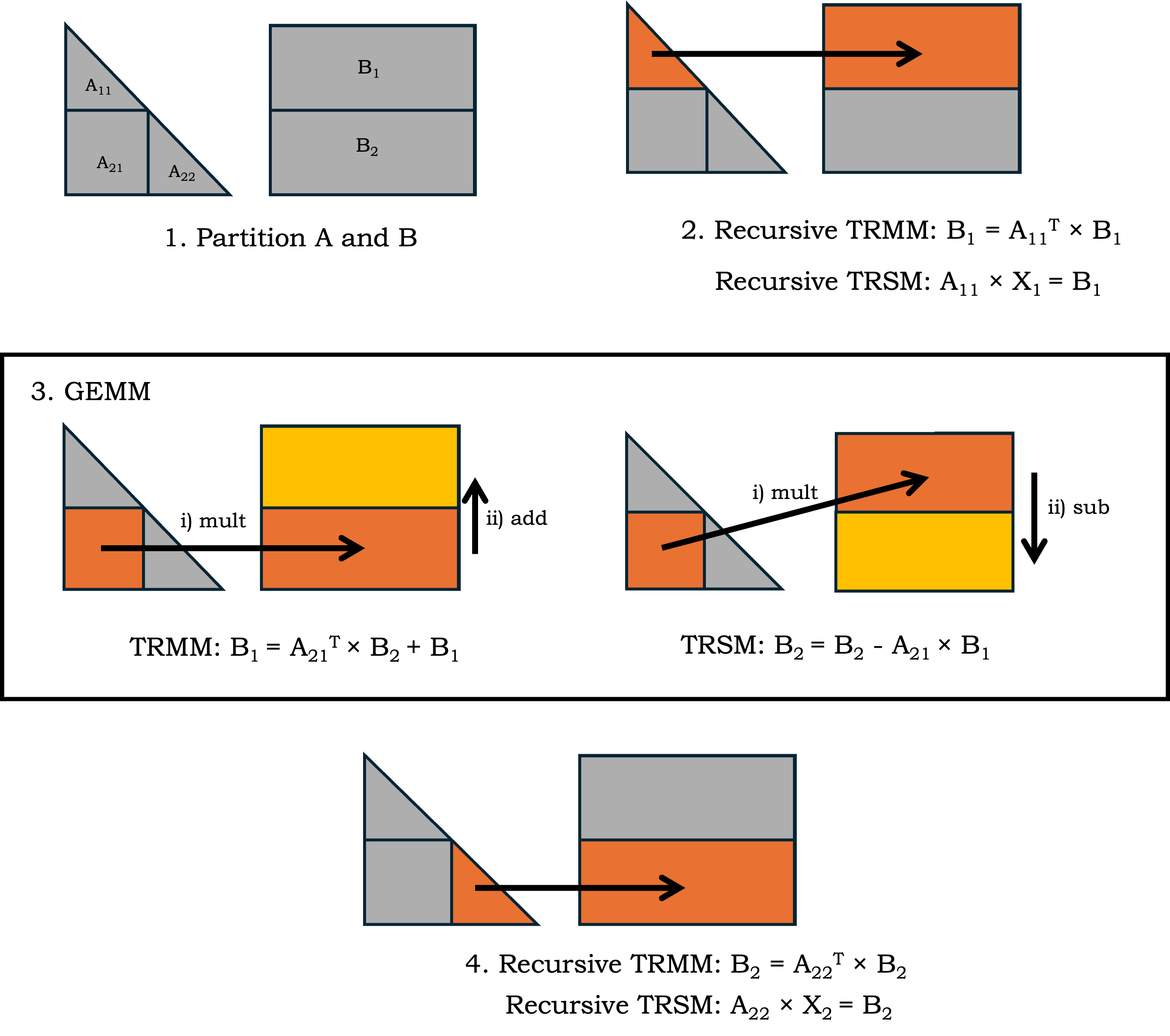}
	\caption{TRMM/TRSM Recursive Illustration.}
	\label{fig:recursive}
\end{figure}

The recursive framework for solving triangular matrix problems takes advantage of the memory hierarchy of the GPU and maximizes parallelism by leveraging a higher number of GEMM calls, which are highly optimized on modern hardware. 
This involves partitioning the input matrices into submatrices, where the triangular matrix  $A$  is divided into a top-left triangular block  $A_{11}$ , a lower-left block  $A_{21}$ , and a bottom-right triangular block  $A_{22}$ , for lower triangular, while the right-hand side matrix  $B$ is split into corresponding blocks  $B_1$  and  $B_2$. The TRMM and TRSM algorithms rely on this framework, previously proposed in~\cite{charara2016redesigning, charara2017framework}, recursively solving smaller triangular systems and updating the remaining blocks until the submatrices are small enough to handle directly. 

In Figure~\ref{fig:recursive}, we describe two scenarios; left lower transpose TRMM $B = \alpha A^T \cdot B $ and the left lower non-transpose TRSM $A \cdot X = \alpha \cdot B$, to illustrate the recursive approach of the operations, where $A$ is an $N \times N$ lower-triangular matrix and $B$ is an $N \times M$ rectangular matrix. In the first step of both TRMM and TRSM, the matrices A and B are subdivided into sub-matrices $A_{11}, A_{21}, A_{22}$, and $B_{1}, B_{2}$ respectively. Using the standard sub-matrix notation, $A_{11}$ refers to the triangular submatrix  corresponding to the first half of $A$, where the block size is chosen as $n = N/2$ to maximize the number of GEMM calls in later stages, thereby improving computational efficiency.  From this partitioning, the TRMM and TRSM operations proceed in three main steps. First, a recursive step: TRMM computes $B_1 = A_{11}^T \cdot B_1$, while TRSM solves $B_1 = A_{11} \cdot X_1$ recursively.  Second, a GEMM update step: TRMM updates the top block using $B_1 = A_{21}^T \cdot B_2 + B_1$, while TRSM adjusts the solution with $B_2 = B_2 - A_{21} \cdot B_1$. Third, a final recursive step: TRMM recursively computes $B_2 = A_{22}^T \cdot B_2$ , while TRSM solves  $B_2 = A_{22} \cdot X_2$. When the argument size falls below a chosen threshold, the recursive TRMM/TRSM calls are replaced with the appropriate base kernels, terminating the recursion. This recursive framework can be extended to other TRMM and TRSM variants, such as when $A$  is upper triangular or transposed.

\subsection{Generalized Recursive Computation}

We developed a unified recursive framework that generalizes both TRMM and TRSM into a single recursive structure by leveraging Julia’s multiple dispatch to dynamically determine the appropriate function at runtime based on the matrix type, operation mode, and function signature. During recursion, the framework calls the same high-level function, but multiple dispatch selects the correct base kernel based on whether the operation involves matrix multiplication (TRMM) or triangular solve (TRSM), whether the triangular matrix is lower or upper triangular, and whether transposition is involved
while ensuring that the correct computational kernel is applied at each level. 
By structuring the recursion in this way, we not only unify the handling of TRMM and TRSM but also enable seamless extensions to other triangular operations, making the framework highly adaptable for future optimizations and different hardware architectures.




\subsection{Kernel Performance Engineering}

The recurive TRMM and TRSM algorithm benefit from using efficient General Matrix-Matrix Multiplication (GEMM) operations, which are compute-bound and thus well-suited for GPUs, and executing the base TRMM and TRSM algorithmns only on small tiles. To benefit from GEMM performance, the solving time of the base kernels needs to be small relative to the solving time of the much larger GEMM operations.


To optimize the base case GPU kernel performance, several performance engineering techniques were applied, leveraging the GPU's architecture to maximize computational throughput and minimize latency.



\begin{itemize}
    \item \textbf{Memory optimization.} We leverage shared memory and contiguous memory striding.
    \item \textbf{Reformulation of algorithm for parallelism.} For the TRSM algorithm specifically, there is an interdependency between rows but not between columns. Parallelism is achieved by distributing row computations across threads for each column. To facilitate this formulation, the algorithm is reformulated so that synchronization occurs after each row.
\end{itemize}

\section{Performance Results}
\label{sec:res}

In order to demonstrate the performance of the unified implementation, this section presents benchmarking of the recursive TRMM and TRSM function on several different types of hardware, and shows its runtime is comparable with state-of-the-art cuBLAS/rocBLAS libraries.

\subsection{Hardware-unified performance}
\begin{figure}
    \centering
    \includegraphics[width=1\linewidth]{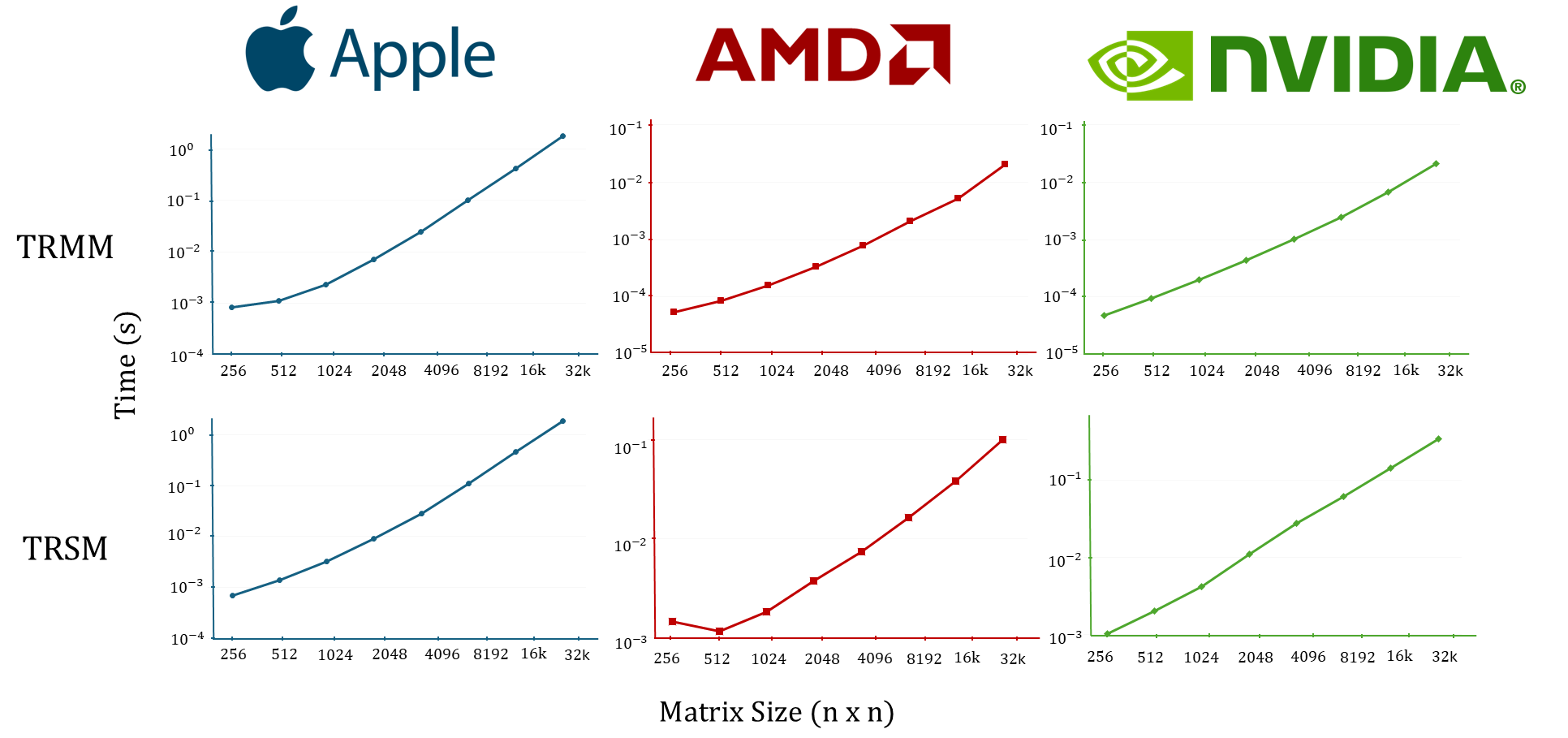}
    \caption{Runtime of recursive unified TRMM (top row) and TRSM (bottom row) functions  across different GPU hardware platforms (Apple, AMD, NVIDIA) as a function of the size of the matrix  \(A \in \mathbb{R}^{ n \times n}\) for a rectangular matrix \(B \in \mathbb{R}^{ n \times 256}\), both of single precision. The figure shows a similar performance trend across hardware, demonstrating similar performance trends on three different hardware setups. }
    \label{fig:portability}
\end{figure}

Figure \ref{fig:portability} shows the runtime performance of TRMM (top row) and TRSM (bottom row) in three different types of hardware, Apple, AMD and NVIDIA GPUs respectively. We can see in the figure that all hardware follows the same performance trend. This is the first time recursive TRMM and TRSM functions are made available for Apple Silicon GPUs and that such a performant hardware-agnostic implementation is made available.

\subsection{Performance versus standard libraries}
\begin{figure}
    \centering
    \includegraphics[width=1\linewidth]{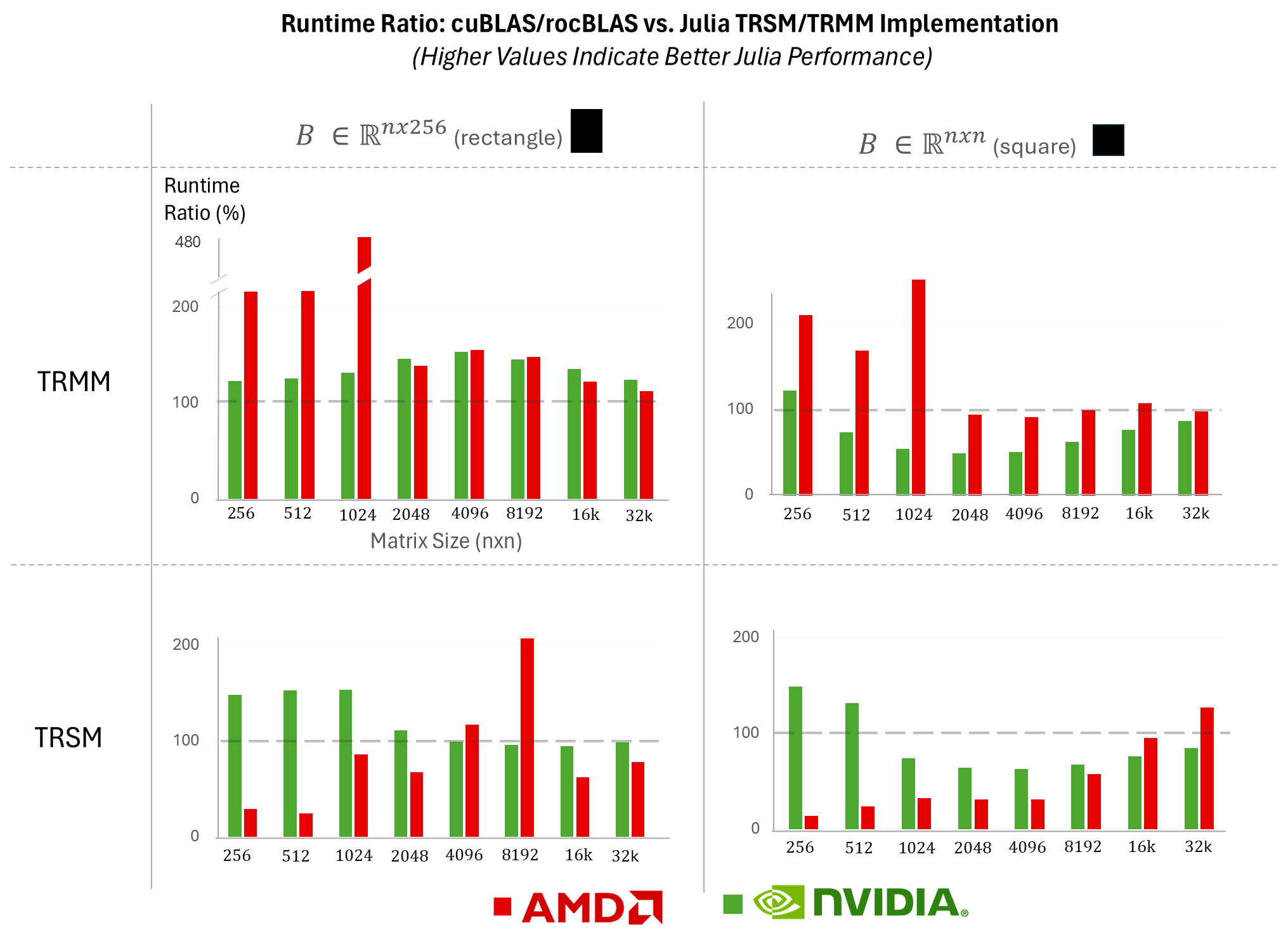}
    \caption{Runtime ratio of cuBLAS/rocBLAS versus the Julia implementation of TRMM (top row) and TRMM (bottom row) in function of the size of matrix \(A \in \mathbb{R}^{ n \times n}\) . Higher values indicate that the Julia implementation is faster, 100\% indicates equal performance. The left two figures are for a matrix \(\in \mathbb{R}^{ n \times 256}\) having a set width. The right two figures show the case of a square matrix \( B \in \mathbb{R}^{ n \times n} \). The figures demonstrates the unified implementation generally performs on par with state-of-the-art specific optimized cuBLAS/rocBLAS libraries. }
    \label{fig:runtime_ratio}
\end{figure}

Figure~\ref{fig:runtime_ratio} shows the ratio of the runtime performance of cuBLAS/rocBLAS versus the runtime performance of Julia TRMM (top row) /TRSM (bottom row) functions for rectangular matrices (left) and and square matrices (right). When the bars exceed the 100\% dashed line, Julia is faster than the respective library.

The TRMM Julia implementation for the rectangular cases (top left figure) consistently outperforms both cuBLAS and rocBLAS. The TRMM Julia implementation for the square case (top right figure) performs similarly to or better than rocBLAS, never falling below 90\% of its runtime, and similarly to cuBLAS at 50\%-200\%. Thus, the unified TRMM is consistently as performant as state-of-the-art libraries.

The TRSM Julia implementation for rectangular cases(bottom left figure) matches or surpasses cuBLAS performance in most cases. For smaller matrices, rocBLAS is faster, but at matrix sizes above 1000 Julia matches at least 2/3 of rocBLAS performance. The Julia TRSM implementation for the square case (right figure) is mostly on par with cuBLAS performance, matching at least 2/3. The Julia implementation matches rocBLAS as matrix size increases, but underperforms at small matrix sizes. It is worth noting that the small matrix sizes concern running times below $10ms$, where performance differences could arise due to hardware idiosyncracy. Furthermore, we observe that the cuBLAS implementation while initially slower than the unified Julia implementation, appears to scale in line with it, while the rocBLAS implementation appears to scale worse. As such, these differences might be due to algorithmic differences in the base kernel implementation. From the TRSM diagrams, we can conclude that at larger matrix sizes and relevent runtimes the unified implementation is on par with both cuBLAS and rocBLAS.

In summary, the results demonstrate that the Julia implementation of TRSM and TRMM is highly competitive with state-of-the-art libraries like cuBLAS and rocBLAS. TRMM shows particularly strong performance, with Julia more consistently achieving or exceeding the performance of rocBLAS and remaining close to cuBLAS. TRSM results indicate the unified implementation performing on par with cuBLAS/rocBLAS at larger matrix sizes where runtime becomes relevant. These findings position the Julia functions as a viable alternative for many computational scenarios, especially where scalability is critical.

The benchmarking demonstrates that the performance of the hardware-agnostic generic implementation for TRSM/TRMM is in line with the performance of specialized state-of-the-art libraries, and that performance-portability is possible with only a few hundred lines of code.



\subsection{Hardware details}

 The experiments utilized the following computing platforms:
\begin{itemize}
  \item \textbf{NVIDIA GPU:} Platform  I consists of a single compute node, including two 28-core Intel(R) Xeon(R) Gold 6330 CPU running at 2.00 GHz, 1008 GB of memory, and four NVIDIA A100 with 80GB GPUs.
  \item \textbf{AMD GPU:} Platform  II consists of a single compute node, including two 64-core AMD EPYC 7713 CPU running at 2.00 GHz, 256 GB of memory, and one AMD MI100 with 32GB GPUs.
  \item \textbf{Apple GPU:} Platform  III consists of a single compute node, including M1 Pro with 10-core CPU, and 16-core GPU and 16 GB of memory.
  \end{itemize}

  We have confirmed the trends of the benchmarking on consumer GPUs as well.

\section{Conclusion}
\label{sec:con}

In this work, we have developed hardware-agnostic implementations of recursive TRMM and TRSM that cover most hardware platforms with a single API using only a few hundred lines of code. The implementation matches the performance of state-of-the-art implementations (cuBLAS/rocBLAS) and for the first makes linear algebra implementations available for Apple Silicon. 

Comparable performance of TRMM and TRSM to both cuBLAS and rocBLAS was found for larger matrix sizes. 
For Apple Silicion, the performance trend is in line with AMD and NVIDIA devices, and expected to be consistent with linear algebra libraries for Apple Silicon that might be developed in the future.

Future work involves advanced scheduling and extension to multi-core hardware settings using
Dagger.jl~\cite{alomairydynamic} to allow users to run high-performance implementations of TRMM and TRSM, relying on software to optimize for the hardware without user effort.
Our results indicate that the Julia Array abstractions and KernelAbstractions provide a performance portable solution for various hardware with minimal code complexity. 


\section*{Acknowledgment}

We would like to acknowledge the work and support of members of the Julia Lab, in particular Valentin Churavy and Julian Samaroo, who both helped with our understanding of available tools within the Julia ecosystem for performance portability.
This material is based upon work supported by the U.S. National Science Foundation 
(awards CNS-2346520, PHY-2028125, RISE-2425761, DMS-2325184, OAC-2103804, and OSI-2029670), the Defense Advanced Research Projects Agency (DARPA HR00112490488), the Department of Energy, National Nuclear Security Administration (DE-NA0003965) and by the United States Air Force Research Laboratory (FA8750-19-2-1000).  Neither the United States Government nor any agency thereof, nor any of their employees, makes any warranty, express or implied, or assumes any legal liability or responsibility for the accuracy, completeness, or usefulness of any information, apparatus, product, or process disclosed, or represents that its use would not infringe privately owned rights. Reference herein to any specific commercial product, process, or service by trade name, trademark, manufacturer, or otherwise does not necessarily constitute or imply its endorsement, recommendation, or favoring by the United States Government or any agency thereof. The views, conclusions and opinions of authors expressed herein  are those of the authors and do not state or reflect those of the United States Government or any agency thereof. They should not be interpreted as representing the official policies, either expressed or implied, of the United States Air Force or the U.S. Government.
We would also like to acknowledge the Belgian American Educational Foundation for financial support for Evelyne Ringoot. In addition, we would like to acknowledge KAUST Ibn Rushd post-doctoral fellowship who supported Rabab Alomairy. We would like to acknowledge IBEX at the Supercomputing Laboratory of the King Abdullah University of Science and Technology (KAUST) in Thuwal, Saudi Arabia for computational resources. 

%
%
%
%
\bibliography{References}

\end{document}